# Calibration of the NEVOD-EAS array for detection of extensive air showers

M.B. Amelchakov[1], A.G. Bogdanov[1], A. Chiavassa[2,3], A.N. Dmitrieva[1], D.M. Gromushkin[1], E.P. Khomchuk[1], S.S. Khokhlov[1], R.P. Kokoulin[1], K.G. Kompaniets[1], A.Yu. Konovalova[1], G. Mannocchi[4], K.R. Nugaeva[1,] A.A. Petrukhin[1], I.A. Shulzhenko[1], G. Trinchero[3,4*], I.I. Yashin[1], E.A. Yuzhakova[1]

[1] *National Research Nuclear University MEPhI (Moscow Engineering Physics Institute), 115409 Moscow, Russia*
[2] *Dipartimento di Fisica dell' Università degli Studi di Torino, 10125 Torino, Italy*
[3] *Sezione di Torino dell' Istituto Nazionale di Fisica Nucleare – INFN, 10125 Torino, Italy*
[4] *Osservatorio Astrofisico di Torino – INAF, 10025 Torino, Italy*

*Corresponding author E-mail address: andrea.chiavassa@unito.it

## Abstract

In this paper we discuss the calibration of the NEVOD-EAS array which is a part of the Experimental Complex NEVOD, as well as the results of studying the response features of its scintillation detectors. We present the results of the detectors energy calibration, performed by comparing their response to different types of particles obtained experimentally and simulated with the Geant4 software package, as well as of the measurements of their timing resolution. We also discuss the results of studies of the light collection non-uniformity of the NEVOD-EAS detectors and of the accuracy of air-shower arrival direction reconstruction, which have been performed using other facilities of the Experimental Complex NEVOD: the muon hodoscope URAGAN and the coordinate-tracking detector DECOR.

## 1. Introduction

Detection of extensive air showers (EAS) is the only way to study primary cosmic rays with energies above $10^{15}$ eV. In most experiments measuring extensive air showers, the reconstruction of the primary particle characteristics is based on the analysis of the energy deposit of shower particles in an array of detectors. Most of energy deposit is due to the interaction of air-shower electromagnetic component with the detector material. Air-shower parameters are estimated by approximating the energy deposits observed in the array of detectors with the lateral distribution function of EAS electrons [1]. In this regard, an important aspect of the experiment is the detector energy calibration, i.e. the determination of the relationship between the responses of the detectors and the energy released in them by EAS particles.

The measured amplitude or charge spectra of detector responses are compared with those obtained by calculations or simulation. In this way, the detectors of the EAS-TOP [2], KASCADE [3], KASCADE-Grande [4], Pierre Auger [5], Tibet air-shower array [6], Ice-Top [7], LHAASO [8], etc. are calibrated.

When an air-shower array is operated together with other facilities (for example, in a complex), it becomes possible to carry out various cross-calibrations. For instance, the EAS parameters reconstructed by Cherenkov water detectors and using the fluorescence method were compared at the Pierre Auger Observatory [9]; the characteristics of EASs recorded simultaneously in the KASCADE and KASCADE-Grande arrays [4] were compared; the air-shower events in the Tunka-133 experiment were used to calibrate the radio method [10]; the IceTop array was used to check the accuracy of muon bundle arrival direction reconstruction in the IceCube detector [11].

In this paper we discuss the calibration of the NEVOD-EAS array included in the Experimental Complex NEVOD [12], as well as the results of a detailed study of the performances of the scintillation detectors used.

## 2. The NEVOD-EAS air-shower array

The NEVOD-EAS array is designed to detect the electron-photon component of extensive air showers with energies in the range from $10^{15}$ to $10^{17}$ eV. It includes 36 scintillation detector stations (DSs) deployed on an area of about $10^4$ m$^2$ around the Experimental Complex NEVOD on the roofs of laboratory buildings and on the ground surface. Stations are combined into 9 clusters. The layout of the DSs and clusters of the array is shown in Figure 1 [12].

Each cluster consists of 4 DSs (Figure 2) located at the vertices of a quadrilateral (mainly a rectangle) with typical side lengths of about 15 m, as well as of a local post of preliminary data processing. The cluster local post receives and digitizes analog signals from detector stations, selects events according to intra-cluster trigger conditions, assigns timestamps to events, and, thus, operates as an independent air-shower installation measuring both the number of particles detected by each DS and the EAS arrival direction.

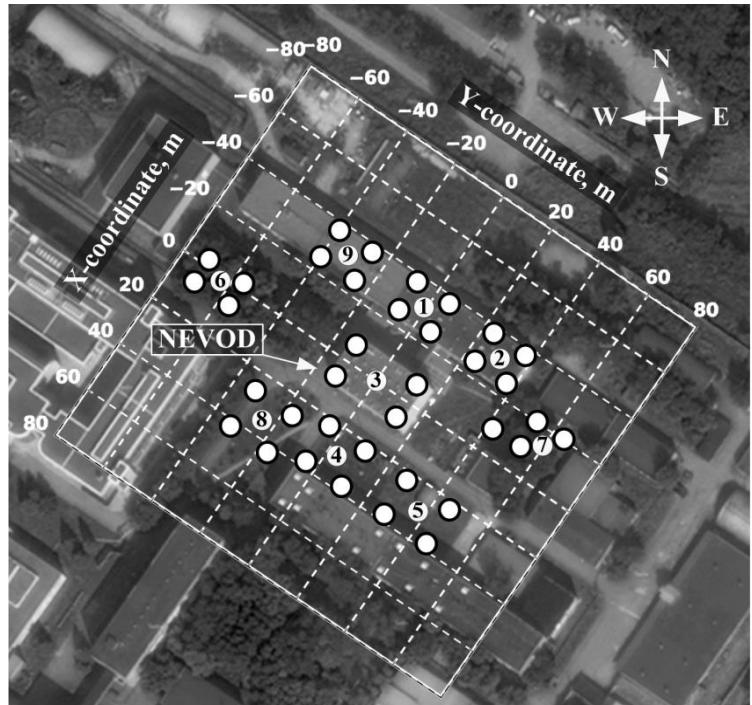

*Figure 1 – The layout of the detector stations and clusters of the NEVOD-EAS array*

The main elements of the NEVOD-EAS array are scintillation detectors of the charged particles, mainly electrons, of extensive air showers (Figure 2). These detectors were previously used in the EAS-TOP [14] and KASCADE-Grande [4] experiments. The NEVOD-EAS detectors [15] consist of a plastic scintillator NE102A with dimensions of 800×800×40 mm$^3$ and one or two photomultipliers (PMTs) Philips XP3462. The scintillator and PMTs are enclosed inside a light-insulated stainless steel pyramidal housing. To improve light collection, the inner surface of the housing is painted with a diffusely reflective coating. The distance between the PMT photocathode and the scintillator is 30 cm.

Each detector station consists of four scintillation detectors installed inside a protective external housing (Figure 2) and has an area of 2.56 m$^2$. Three detectors of the station are equipped with one PMT operating at high gain (in the following named "standard"). The fourth detector includes two PMTs: the standard photomultiplier and an additional one working at a lower gain. Standard PMTs are used to measure the EAS particle densities of up to ~ 100 particles/m$^2$ and for time measurements. The additional PMT expands the DS dynamic range up to ~ 10$^4$ particles/m$^2$ when detecting EASs with high particle density.

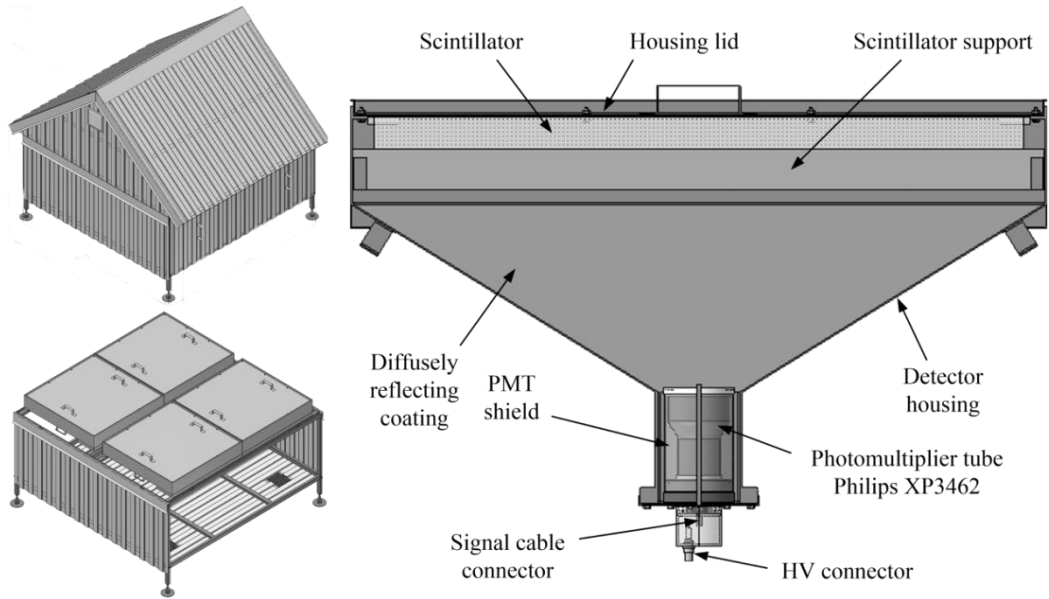

*Figure 2 – The design of the NEVOD-EAS detector station: top left – detector station, bottom left – location of detectors inside the DS, right – scintillation detector*

### 3. Simulation of the NEVOD-EAS detector and detector station

To perform the energy calibration of the NEVOD-EAS detector station, we have developed the models of the scintillation detector and DS using the Geant4 software package [16]. The models take into account the geometry of the detector and DS, as well as physical and chemical properties of materials and surfaces.

The detector geometry includes the following main elements: a metal housing, a scintillation plate, two scintillator's supports (see Figure 2), glass and a photocathode of PMT.

Sheet steel with thickness of 1 mm is used as the housing material. The reflection coefficient, of the coating used to paint the inner surface of detector housing, is 0.9.

The NE102A scintillator [17] has the following properties: a density of 1.032 g/cm$^3$, a light refractive index of 1.58, a light yield of 12000 photons/MeV, an emission time of 2.4 ns, and a light absorption length of 1 m. The dependence of the scintillator relative light yield on the energy of emitted photons, which we embedded in the model, is shown in Figure 3 (curve No. 1).

The scintillator's supports have the shape of parallelepipeds and consist of organic glass with a density of 1.19 g/cm$^3$, a refractive index of 1.49, and a light attenuation length of 5 m.

The input window of the PMT represents a cylinder with a diameter of 76 mm and a height of 13 mm. The material of the input window is glass with a density of 2.53 g/cm$^3$, a refractive index of 1.54, and a light attenuation length of 5 m. The hemispherical PMT photocathode is set as a metal surface.

The probability of photoelectron emission was simulated using the dependence of the photomultiplier quantum efficiency on the photon energy, which is shown in Figure 3 (curve No. 2). The signal charge at the PMT output was modeled as the sum of PMT responses to single photoelectrons. The charge of photomultiplier response to each photoelectron was drawn according to the normal distribution, the parameters of which were determined from the peak of a typical charge spectrum of PMT Philips XP3462 responses to single photoelectron illumination, measured at a gain of $2 \times 10^6$ (Figure 4). The average charge of single-electron signals is $0.31 \pm 0.01$ pC, and the FWHM of the distribution is ~ 0.3 pC.

The model of the detector station includes an array of four identical scintillation detectors placed inside an external housing made of sheet steel with a thickness of 0.7 mm. When simulating the passage of particles, the response of the detector station is the sum of the responses of 4 scintillation detectors included in it.

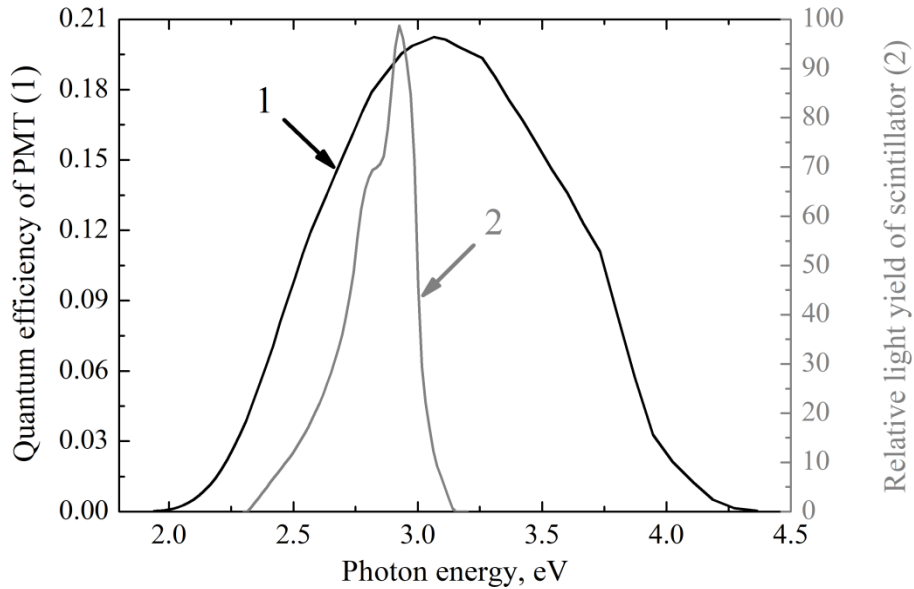

*Figure 3 –The quantum efficiency of PMT [18] and relative light yield of scintillator [17] as functions of photon energy*

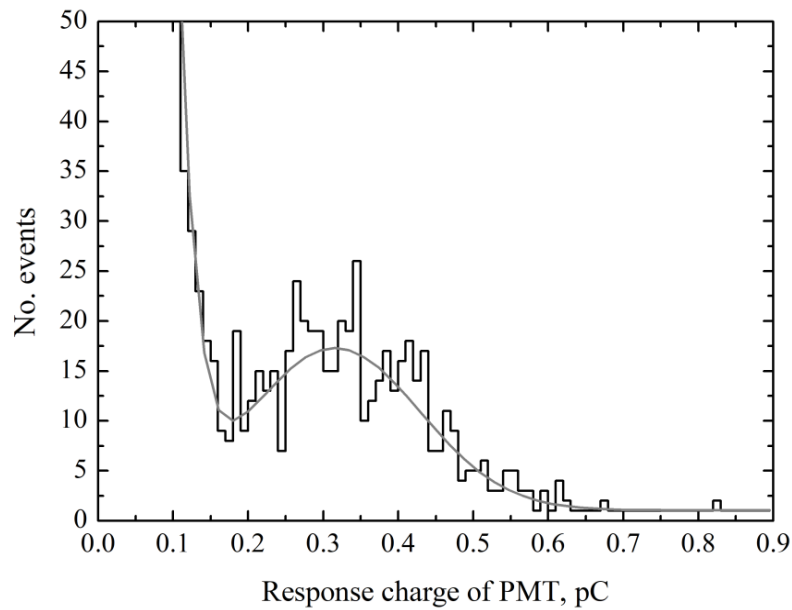

*Figure 4 – Charge spectrum of PMT responses to single-electron illumination (histogram – experimental distribution, curve – approximation by the sum of two normal distributions)*

## 4. Verification of the detector model

To verify the model of scintillation detector, we have compared the non-uniformity of the detector response (dependence of the detector response on the place where the charged particle passes through the scintillator) obtained experimentally and as a result of the simulation.

An experimental study of the non-uniformity of light collection was carried out using the supermodule (SM) of the muon hodoscope URAGAN [19]. The URAGAN supermodule is a coordinate-tracking detector with an area of 11.5 m². It consists of eight coordinate planes based on gas discharge tubes operating in a limited streamer mode. With the supemodule single muon tracks are reconstructed with high spatial and angular accuracy (1 cm and 0.8º, respectively) in the zenith angle range from 0º to 80º.

The facility to measure the non-uniformity of light collection (Figure 5) consists of the SM URAGAN, on the surface of which the understudied detector is installed, a digital oscilloscope Tektronix MDO3034 recording signals from the detector, and two personal computers (PCs). One PC ensures operation of the SM, the second one receives data from the oscilloscope and combines them with the data of supermodule. When a charged particle passes through the SM, the supermodule generates a trigger for the oscilloscope. In its turn, the oscilloscope transmits the digitized waveform of the recorded detector signal to the PC of the facility. At the same time, this computer receives information about the particle track coordinates and direction form the PC of the supermodule. For the analysis, we selected only events with single, almost vertical muons crossing the scintillator at zenith angles smaller than 15°.

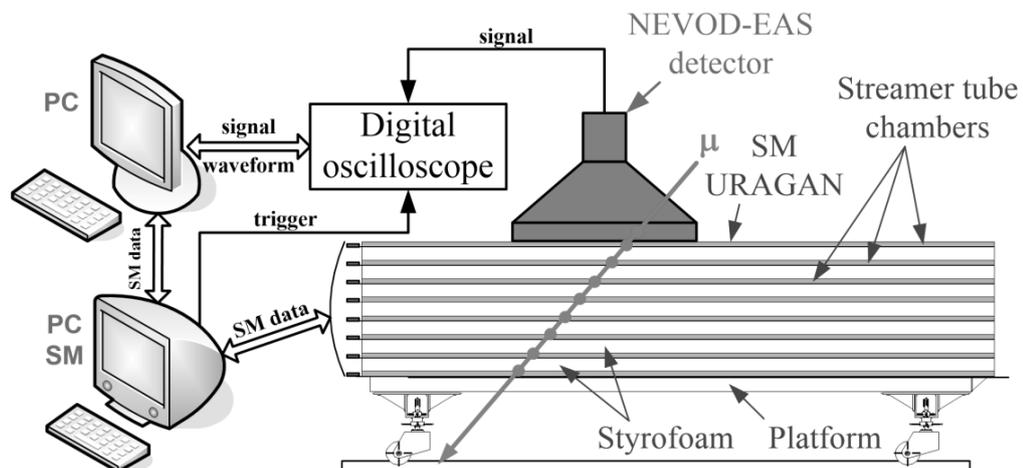

*Figure 5 – Scheme of the facility for measuring light collection non-uniformity of detectors [15]*

Figure 6 (left) shows the experimentally measured matrix of average charges of the detector responses to the passage of single muons. The matrix cell size is 1×1 cm². The statistical reliability of each cell is ~ 15 events. The coordinates of the cells in centimeters are plotted along the matrix axes. The top and side graphs show the variation of the detector response charge relative to the charge, averaged over the scintillator area, in two mutually perpendicular sections. The sections are marked on the matrices with dashed lines. As can be seen, the maximum light collection is observed from the area which in tests is located directly under the PMT (in the experiment it is located directly above the PMT). The observed shift of this area relative to the center of the detector is due to the non-central location of the PMT mounts in the housing due to the need to install an additional photodetector.

A similar matrix was obtained by simulating the detector response to single muons passing through different parts of the scintillator surface at zenith angles smaller than 15° (Figure 6, right). The simulation of muons was carried out taking into account their spectrum which was calculated using the CORSIKA [20] (description is given in the next section).

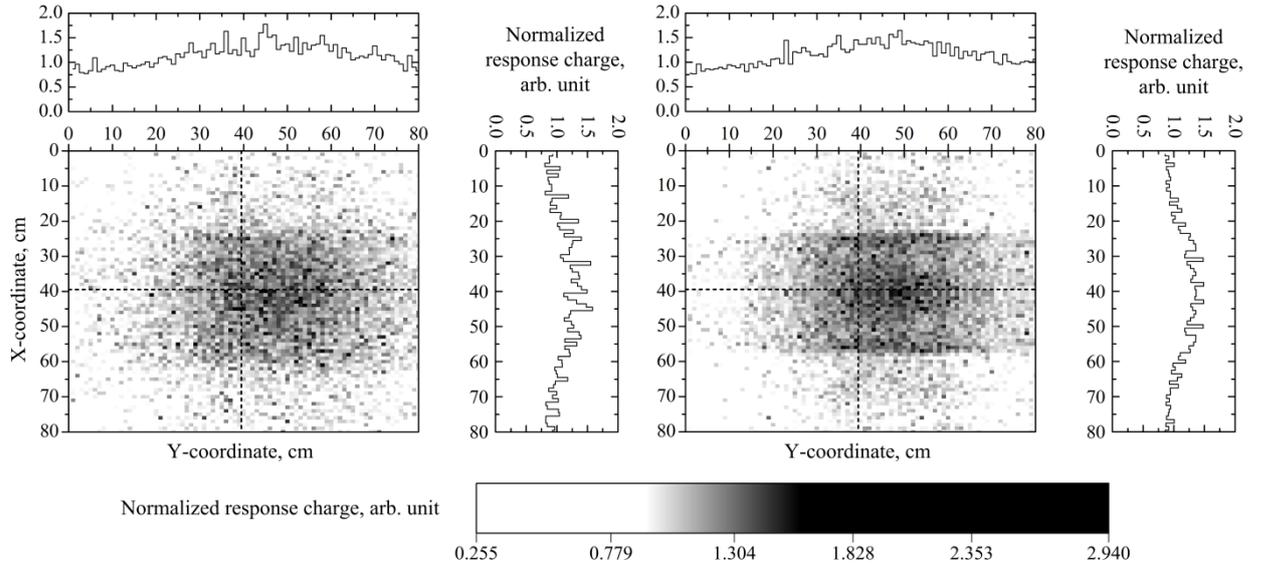

*Figure 6 – Matrices of the normalized response charge of the NEVOD-EAS detector obtained experimentally (left) and from results of simulation (right)*

As seen from Figure 6, the shapes of the experimental and simulated matrices are in good agreement. In both cases, the light collection non-uniformity is defined as the ratio of the standard deviation calculated from the array of all cells to the average value. The experimental value of the light collection non-uniformity is 18.4 ± 0.1%, while the simulated value is 18.2 ± 0.1%. Thus, it can be concluded that the developed model gives a response close to the response of a real NEVOD-EAS scintillation detector, and can be used for its energy calibration.

## 5. Energy calibration of the detector station

During the operation of the NEVOD-EAS array, the monitoring is performed every 4 hours. In the monitoring, the charge spectra of all 36 detector stations are measured in the self-triggering mode. In this mode, signals from detector stations are mainly due to the passage of single muons. To a lesser extent, the signals can be caused by the passage of hadrons, high-energy electrons and gammas. The typical most probable charge of the muon peak of the NEVOD-EAS detector stations is about 13 pC [13]. For the energy calibration of the DSs, it is necessary to determine the energy deposit corresponding to the peak value of the charge spectrum obtained in monitoring mode.

Using the developed model, we have simulated the DS response to single particles: muons, electrons, protons and gammas. When drawing the tracks of muons, protons and gammas, we used differential spectra for different values of the zenith angle [21] obtained by simulation with CORSIKA [20]. To simulate the energy of electrons and gamma-rays, we used a differential spectrum combining the results of CORSIKA simulations for kinetic energies above 100 MeV and those of calculation [22] for kinetic energies below 100 MeV. An example of the spectra of the abovementioned particles for the vertical direction is shown in Figure 7.

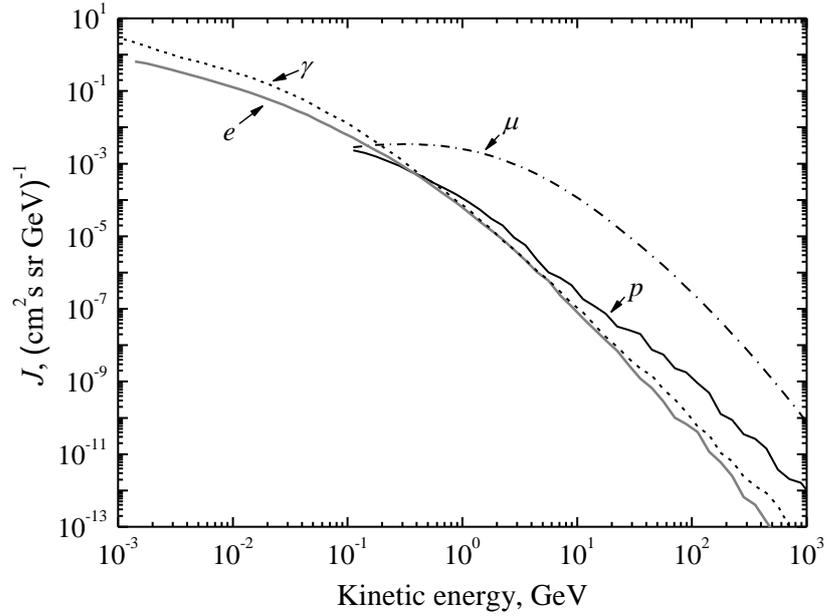

*Figure 7 – The example of the energy spectra of particles used in simulation (muons, electrons, protons and gammas) for vertical direction*

Figure 8 shows the distributions of the energy deposit of single particles inside the NEVOD-EAS detector station calculated by the simulation: for electrons, for muons, for gammas and summed for all types of particles. There are two peaks in the presented distribution. The left peak is due to the energy losses of electrons and gammas with energies of several MeV. The right one with the most probable value $\delta E_{peak}$ of 11.5 MeV and the full width (FWHM) of 5.3 MeV is mainly contributed by the energy deposits of muons and electrons. At values greater than 8.6 MeV (0.75 of the most probable value $\delta E_{peak}$), the average energy deposit of muons is 14.8 MeV, and their contribution to the summed distribution is 71.6%. The average energy deposit of electrons is 14.4 MeV with a contribution of 22.8% to the summed distribution. Protons (1.9%) and gammas (3.7%) with average energy deposits of 34.9 and 17.8 MeV, respectively, make small contribution to the summed distribution. Due to the insignificance of their contributions, individual distribution for protons is not presented. The average energy deposit of all particles in DS $\langle\delta E\rangle$ is 15.2 MeV.

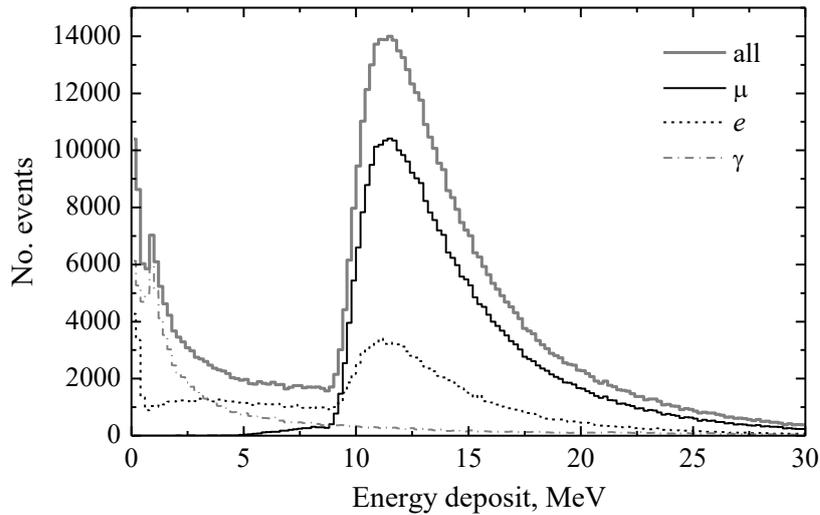

*Figure 8 – Distributions of the energy deposit of single particles in the scintillator (results of simulation using the Geant4 software package)*

Next, we have compared the charge spectrum of the DS responses obtained from simulated events with the experimental charge spectrum measured in self-triggering mode. The experimental and simulated spectra are shown in Figure 9 as a histogram and a curve, respectively. The left peak in the experimental distribution is explained by the contribution of the natural radioactivity background and of the PMT dark noises. It is quite difficult to take these factors into account in simulation, so the left peak is absent in the spectrum of simulated DS responses. The right peak in the experimental distribution is due to the passage of cosmic ray particles. Its shape is well reproduced by the spectrum of simulated DS responses. The most probable response charge $Q_{peak}$ is 13 pC, the full width at half maximum of the distribution $FWHM_Q$ is 9.1 pC.

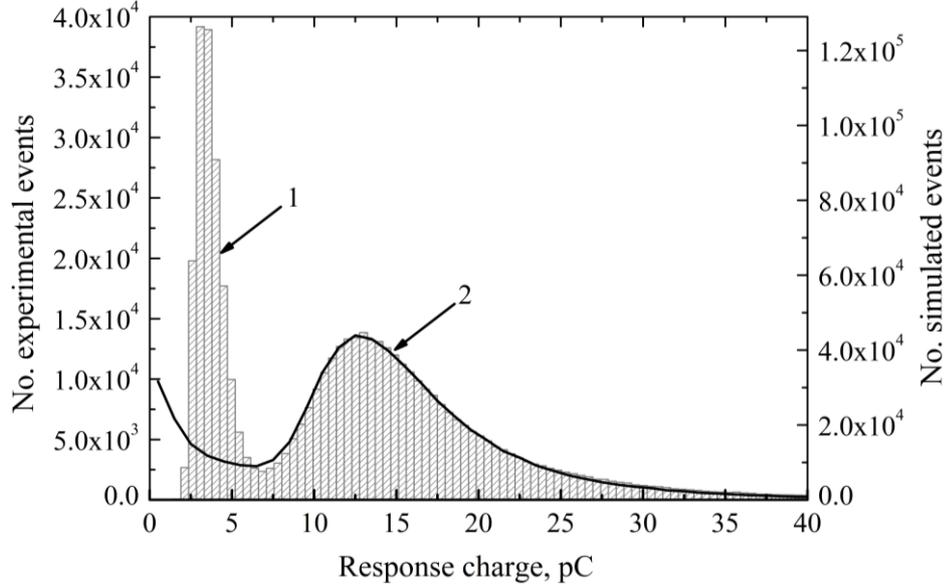

*Figure 9 – Experimental (histogram "1") and simulated (curve "2") charge spectra of the responses of the NEVOD-EAS detector station obtained in self-triggering mode*

It should be noted that the relative peak width of the charge spectrum ($FWHM_Q/Q_{peak} = 0.7$) is noticeably larger than the relative width of the energy deposit peak ($FWHM_{\delta E}/\delta E_{peak} = 0.46$). This is explained by additional fluctuations, which are introduced into the DS response by the light collection non-uniformity of the detector and by the Poisson fluctuations of the processes in the PMT. At a light collection non-uniformity of 18.1%, the contribution to the relative peak width can be estimated as ~2.35×0.18 = 0.42, i.e. the contribution of the light collection non-uniformity is comparable with energy deposit fluctuations. The contribution of Poisson fluctuations in the PMT can be estimated from the number of photoelectrons: at the most probable response of 13 pC the PMT detects ~ 42 photoelectrons. Thus, the contribution of Poisson fluctuations is ~2.35/$\sqrt{42}$= 0.36. The combination of energy deposit fluctuations, light collection non-uniformity and Poisson fluctuations in the PMT determines the relative width of the charge distribution peak ($0.46^2 + 0.42^2 + 0.36^2 \approx 0.7^2$).

The calibration coefficient, necessary for the reconstruction of experimental events, can be calculated as the ratio of the obtained values of $\delta E_{peak}$ and $Q_{peak}$ and is equal to 0.88 MeV/pC.

## 6. Response of scintillation detector to EAS electrons

When reconstructing the parameters of extensive air showers, it is assumed that the main contribution to the measured energy deposit is made by the electron-photon component.

To study the response of the scintillation detector to the EAS electron-photon component using the developed DS model we have simulated its response to electrons and gammas of fixed energies in the range from 10 to 100 MeV. Figure 10 shows dependence of the average energy deposit of vertical electrons and gammas on their energies.

Gammas contribute to energy deposit through two processes: the Compton effect and the production of electron-positron pairs. Interactions of gammas can occur both in the scintillator itself and in the steel housing lid installed above the scintillation plate. According to the database [23], the cross sections of Compton effect in steel and polystyrene are close to each other for gammas with energies from 10 to 100 MeV. At the same time the cross section of pair formation in steel is 3.5–3.8 times higher than in the scintillator. Taking into account the thicknesses of the steel lid (0.8 g/cm$^2$) and the scintillator (4.1 g/cm2), one can obtain that the pair production in the lid and in the scintillator make almost the same contribution to the energy deposit of gammas. For gammas with energies less than 30 MeV, the cross section of Compton effect in polystyrene is higher than those for the pair formation. So in this energy range, the Compton effect in polystyrene causes up to 50% of the energy deposit of gammas. It should be noted that, in this case, the Compton effect in the lid makes almost no contribution due to the small probability of interaction. The overall increase in the average energy deposit of gammas in the NEVOD-EAS detector is explained by the growth of the pair production cross section, which is 2.9 times for polystyrene and 2.6 times for steel in the energy range of gammas from 10 to 100 MeV.

As can be seen from the figure, the average energy deposit of particles increases with the growth of energy. As a rule, electrons with energy of 10 MeV lose part of their energy in the steel housing and then stop in the scintillator; therefore, their average energy losses significantly differ from the losses of more energetic electrons. At the critical electron energy (93.1 MeV for the NE102A scintillator) the average losses for electrons are about 11.5 MeV. Since the losses of the vertical minimum ionizing particle in the scintillation detector are 8.0 MeV, the energy deposit of electrons with critical energy is equivalent to the energy deposit of ~1.4 MIP. A weak increase in the average energy deposit of electrons with the energy is explained by the fact that the radiation length of the scintillation plate (41.3 cm) exceeds its thickness (4 cm) by an order of magnitude, and the bremsstrahlung gammas leave the scintillator without interaction.

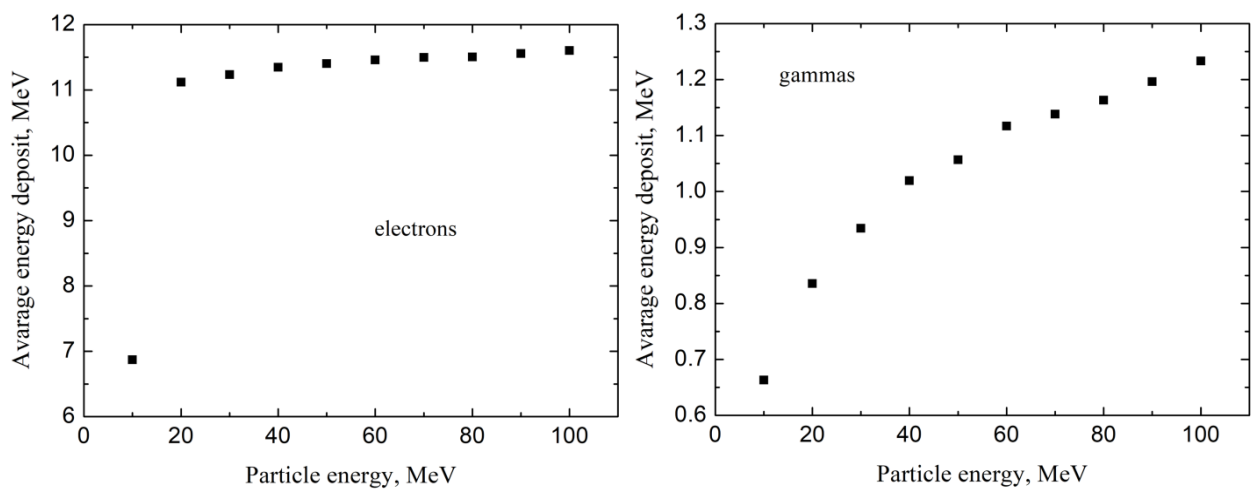

*Figure 10 – The average energy deposit of vertical electrons (left) and gammas (right) in the detector station as a function of their energy*

We have also studied the response of the scintillation detector to electrons and gammas with energies simulated according to the spectrum of EAS particles. To do this, in the CORSIKA program using the models of hadronic interactions QGSJET-II-04 and FLUKA 2020.0.3, we have simulated extensive air showers from protons with energies of $10^{15} - 10^{17}$ eV, distributed by a power-law energy spectrum with the exponent ($\gamma + 1$) of 2.7, and with zenith angles in the range from 0° to 30°. Using the simulation data, we have obtained the energy spectra (Figure 11) of electrons and gammas of air-showers with a size greater than $10^5$ electrons for two cases: for all EAS particles and for particles of EAS central part (within 100 m from the EAS axis); the threshold by the particle kinetic energy was 1 MeV. According to the review [1], the typical energies of electrons and gammas in air-showers are 40 MeV and 10 MeV, respectively. The mean logarithmic energies of particles in the obtained spectra are 24 MeV for electrons and 9 MeV for gammas for the first sample of particles (all EAS particles) and 30 MeV for electrons and 12 MeV for gammas for the second one, i.e. they are in a qualitative agreement with the abovementioned review.

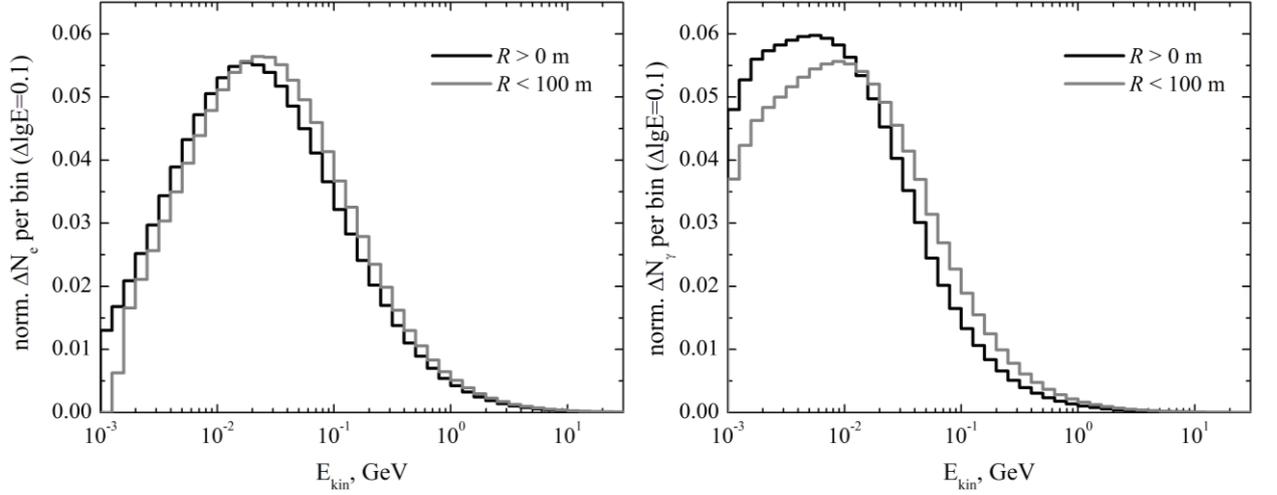

*Figure 11 – The energy spectra of EAS electrons (left) and gammas (right) for two cases: for all EAS particles and for particles within 100 m from the EAS axis*

Using these spectra, we have simulated the response of the NEVOD-EAS detector station to electrons and gammas. Figure 12 shows the distributions of simulated events by the energy deposits of EAS electrons and gammas in the detector station for two samples of particles.

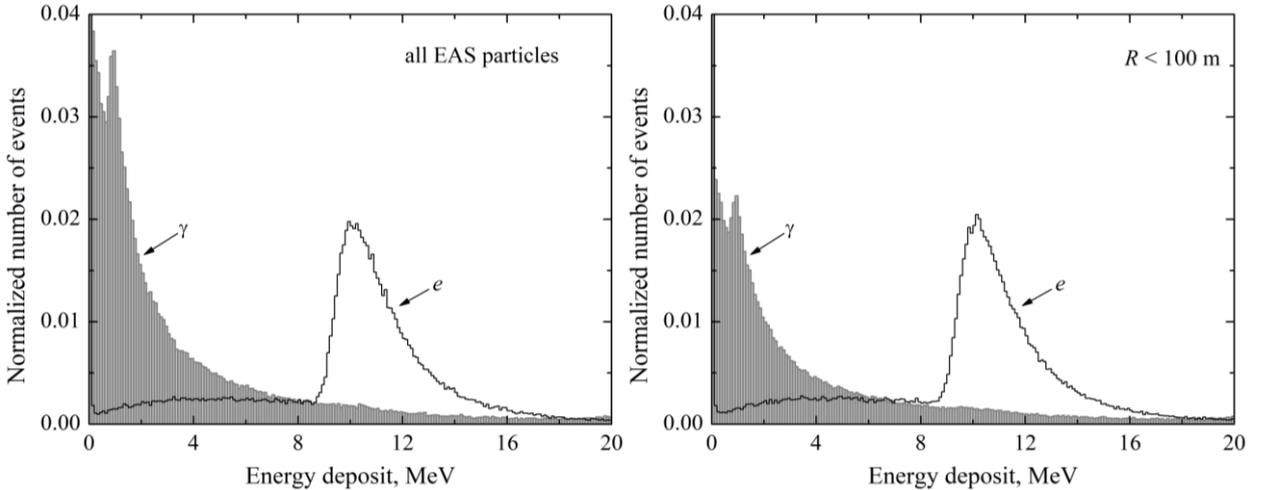

*Figure 12 – The spectra of energy deposits of electrons and gammas in the detector station for two samples of particles: all particles (left) and particles within 100 m from the EAS axis (right)*

Most part of gammas passes through the scintillation detector without interaction. When simulating using the spectrum of all EAS gammas, the energy deposit is observed only in 15.8% of events. When simulating using the spectrum of gammas of EAS central part, the energy deposit is observed in 14.8% of events. In both cases, the spectra of energy deposits of gammas have peaks with the most probable value of 0.95 MeV. The average energy deposits are 0.64 and 0.7 MeV, correspondingly.

Some part of EAS electrons also does not produce energy deposit in the detector, because they stop in its steel lid. The energy deposit in the scintillator is provided by 81.1% of all EAS electrons and 81.7% of electrons from the EAS central part. In the spectra, the peaks with most probable values of 9.95 MeV for all EAS electrons and 10.15 MeV for electrons, arriving at a distance of less than 100 m from the axis, are observed. In both cases, the FWHM of these peaks are 2.5 MeV. The average energy deposit of electrons is 8.22 MeV for the first sample of particles and 8.26 MeV for the second one.

The average energy deposit of electrons from the EAS central part (8.26 MeV) is close to the value used for EAS parameters reconstruction in the EAS-TOP (8.2 MeV [2]) and KASCADE-Grande (8.5 MeV [4]) experiments, where these scintillation detectors were previously operated.

Thus, the obtained values of the average energy deposits of EAS electrons, as well as of the energy calibration coefficient, make it possible to estimate the number of particles that passed through each DS in an EAS event and, based on that, to reconstruct the air-shower size.

### 7. Estimation of the detector station timing resolution

The extensive air shower arrival direction is calculated on event-by-event basis, using the difference of the response times of the cluster detector stations. Therefore, the accuracy of EAS arrival direction reconstruction depends on the timing resolution of the DS.

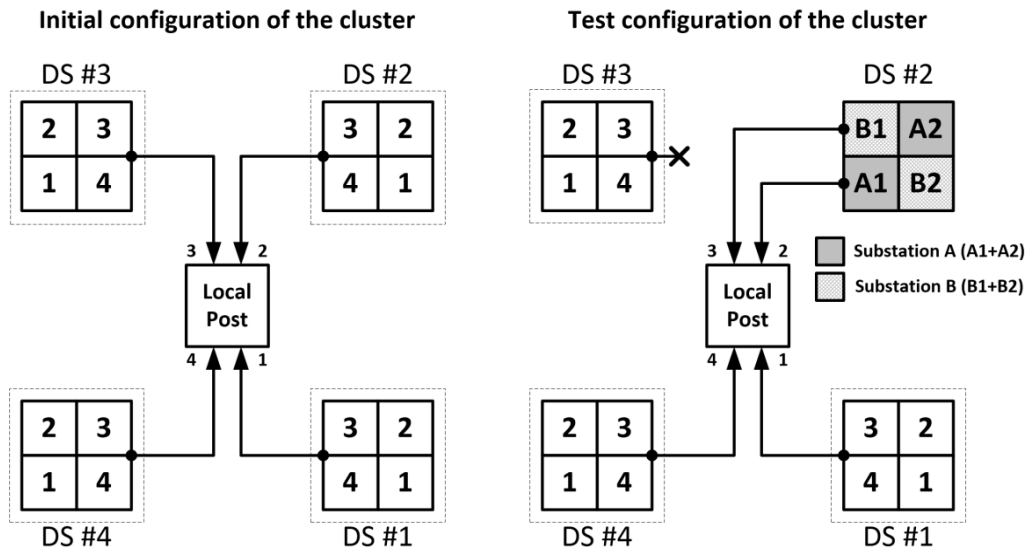

*Figure 13 – Test configuration of a cluster for estimating timing resolution of the NEVOD-EAS detector station*

To estimate the timing resolution of the detector station, we have carried out a test run of data taking at one of the NEVOD-EAS array clusters. The cluster configuration was changed as follows (see Figure 13). One of four cluster DSs was disconnected from the recording system of the local post of preliminary data processing. Scintillation detectors of another station were divided into two substations A and B, consisting of two detectors located on the diagonals of the DS. Substation A remained connected to the same channel of the cluster recording system as the original DS. Substation B was connected to the channel of the previously disconnected DS. In such configuration, when detecting EAS, both substations must respond to the passage of air-shower front almost simultaneously regardless of its the arrival direction. The duration of the test run was about 24 hours. The coincidence of two full DSs and two substations was used as trigger condition. The detection threshold of the cluster measuring channels corresponded to ~ 0.75 of the muon peak most probable value. During the test run, 10519 events were detected.

Figure 14 shows the distribution of events by the differences between the response times of substations A and B. The average value of the distribution is close to zero. That means that in all events the substations responded almost simultaneously. The FWHM is ~ 4.4 ns.

Since the distribution presents the difference between the response times of two substations, the timing resolution of one DS is $\sqrt{2}$ times less.Thus, the error of determination of the DS response time is ~ 3.1 ns. Taking into account that the DS time resolution depends on the PMT jitter, the scintillator emission time, and the ADC sampling frequency, the value obtained from the experimental estimation is quite satisfactory.

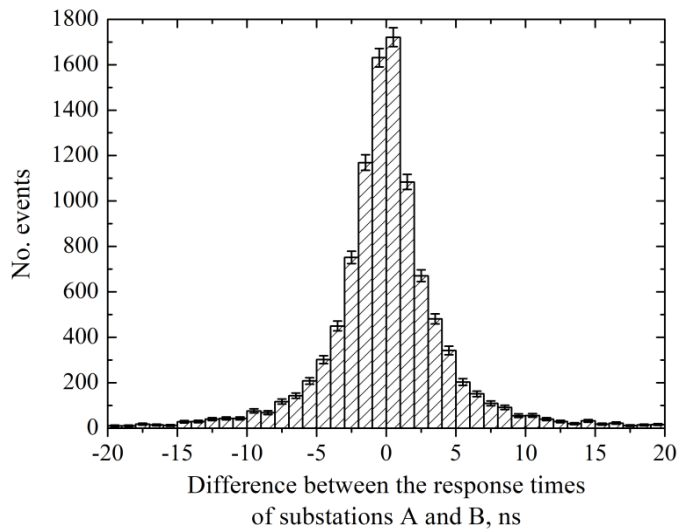

*Figure 14 – Distribution of events by the differences between the response times of substations*

## 8. Accuracy of EAS arrival direction reconstruction

The accuracy of EAS arrival direction reconstruction was determined by comparing the results of reconstructions carried out according to the data of the NEVOD-EAS array and the coordinate-tracking detector DECOR.

The coordinate-tracking detector DECOR is located inside the building of the Experimental Complex NEVOD around the Cherenkov water calorimeter (Figure 15). It consists of 8 vertical supermodules with a total area of 70 m$^2$ [24]. The supermodules have good spatial and angular resolution (~ 1 cm and ~ 1°) for muon tracks. The DECOR detector allows the detection of muon bundles in inclined EASs and the measurement of particle density of bundles [25]. Since muon bundles retain the primary particle direction with good accuracy, the EAS direction reconstructed by the muons can be compared with the direction obtained by the electron-photon component.

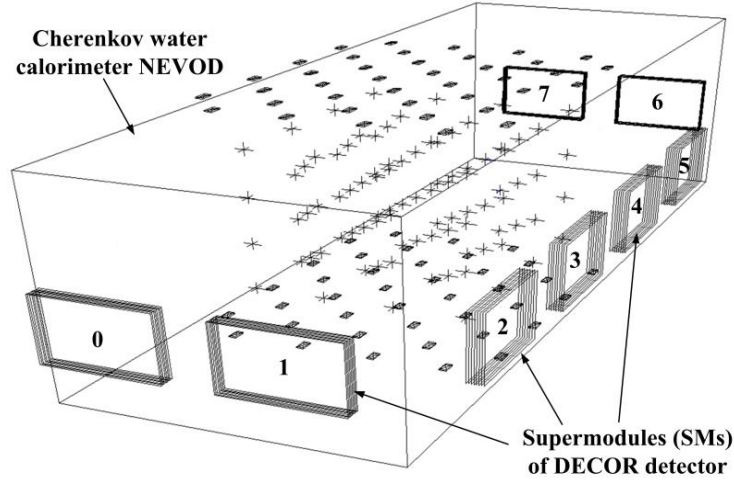

*Figure 15 – Cherenkov water detector NEVOD and coordinate-tracking detector DECOR*

The NEVOD-EAS array and the coordinate-tracking detector DECOR are connected to the global time synchronization system [12] of the Experimental Complex NEVOD, which ensures timestamping of events recorded by these facilities with accuracies of 10 and 25 ns, respectively.

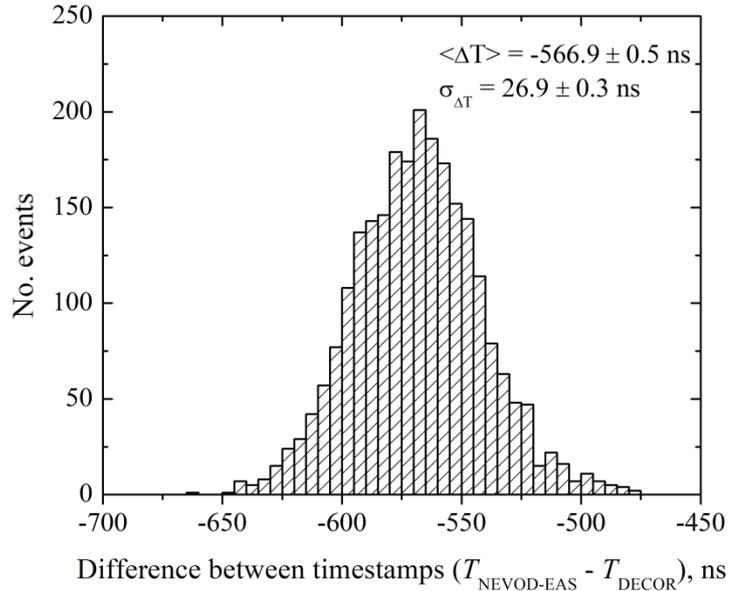

Difference between timestamps ($T_{\text{NEVOD-EAS}} - T_{\text{DECOR}}$), ns

*Figure 16 – Distribution of events in the NEVOD-EAS array and coordinate-tracking detector DECOR by the differences of timestamps*

Figure 16 shows the distribution of events in the NEVOD-EAS and DECOR by the differences of timestamps. It is seen that on average, the DECOR detector is triggered 570 ns later than the NEVOD-EAS, and the vast majority of events falls within the range of ± 100 ns from the average value of the distribution. The observed time shift and width of distribution are due to the operation features of the DECOR trigger system in various classes of events. In further analysis, events in the NEVOD-EAS array and in the DECOR detector were considered as joint ones, if the difference in timestamps was in range from 470 to 670 ns.

For the joint analysis, from the data of the NEVOD-EAS array we selected events in which at least five clusters had been triggered. From the DECOR data, we selected events in which 3 or more muon tracks had been detected by at least 3 supermodules. We have analyzed a 70-day period of operation of the facilities and selected 2456 joint events with arrival direction zenith angles in the range from 10° to 60°. Due to the vertical arrangement of the supermodules, the acceptance of the coordinate-tracking detector DECOR increases with the growth of zenith angle. Thus, 80% of the selected joint events have zenith angles of arrival direction in the range from 20° to 45°.

When analyzing the NEVOD-EAS data, we used two methods to reconstruct the air-shower arrival direction. In the first method, the reconstruction of the direction was preliminary carried out according to the data of individual clusters, and the resulting EAS direction vector was determined as the average vector of all "cluster" directions. In the second, traditional method, we considered all detector stations of the NEVOD-EAS as elements of a single array, and the reconstruction of the air-shower arrival direction was carried out according to the data of all hit DSs. Taking into account the small size of the facility compared to the geometrical sizes of EASs, we used the approximation of a flat air-shower front in both methods.

To estimate the accuracy of EAS arrival direction reconstruction, for each of the selected events we determined the angular deviation between the muon bundle direction obtained with the DECOR and the air-shower direction reconstructed by the NEVOD-EAS data. The resulting distributions of joint events by the angular deviation are shown in Figure 17. Large values of the angular deviation can be associated with the detection of small-sized showers or showers with axes falling on the periphery of the facility.

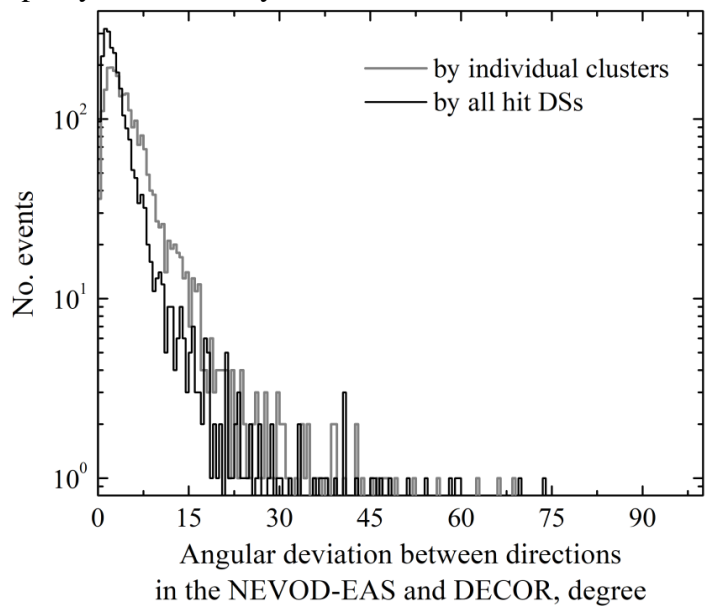

*Figure 17 – Distribution of joint events by angular deviations of directions reconstructed according to the data of the DECOR detector and the NEVOD-EAS array*

We have chosen the boundary, to the left of which fall 68.3% of all events, as an estimate of the arrival direction reconstruction accuracy. When reconstructing according to the data of individual clusters, the accuracy is 5.9°±0.1°. Such accuracy is mainly determined by the temporal resolution of the DS (~3.1 ns), since for a typical cluster with dimensions of 15×15 m$^2$, the expected accuracy of arrival direction reconstruction is ~ 1/15 radian. In [13], using simulated events it was shown that such method for air-shower direction reconstruction provides accuracy better than 5° for primary particle energies above 1 PeV. Thus, the obtained experimental results are in good agreement with the simulation.

When reconstructing the direction according to the data of all hit DSs, the accuracy of cluster synchronization and the deviation of the EAS front shape from the plane becomes the determining factors. The reconstruction accuracy of this method improves up to 3.5°±0.1°.

### 9. Conclusion

For the energy calibration of the detector stations of the NEVOD-EAS air-shower array, a DS model has been developed using the software package Geant4. The model has been verified using experimental data obtained in the study of the non-uniformity of light collection of the NEVOD-EAS scintillation detector carried out at the muon hodoscope URAGAN. A good agreement between the simulation and experimental results has been shown.

Simulating the DS response to single muons, electrons, protons, and gamma rays, we have shown that the most probable value of the charge spectrum peak (typical value is 13 pC), measured at all NEVOD-EAS stations in the monitoring mode, corresponds to energy deposit of 11.5 MeV. Thus, the coefficient for converting the DS response charge into the energy deposit is 0.88 MeV/pC.

Also using the developed DS model, it has been determined that the typical energy deposit of EAS vertical electrons in a detector station is 8.26 MeV.

Comparing the EAS arrival directions reconstructed, in joined events, by the NEVOD-EAS and by the DECOR detectors, we have obtained that EAS arrival direction is measured by the air-shower array with a 3.5-degree accuracy.

### Acknowledgments

The work was performed at the Unique Scientific Facility "Experimental Complex NEVOD" at the expense of the grant from the Russian Science Foundation No. 22-72-10010, https://rscf.ru/project/22-72-10010/.

The authors would also like to express gratitude to the KASCADE-Grande collaboration and the Institut für Astroteilchenphysik of the KIT for their material contribution to the maintenance and development of the NEVOD-EAS air shower array.